\documentclass[12pt,a4paper]{article}

\usepackage[english]{babel}
\usepackage[a4paper,top=2cm,bottom=2cm,left=3.cm,right=2.3cm,marginparwidth=1.75cm]{geometry}
\usepackage{physics}
\usepackage{amsmath}
\usepackage{graphicx}
\usepackage[colorlinks=true, allcolors=blue]{hyperref}

\usepackage[title]{appendix}
\usepackage{mathrsfs}
\usepackage{amsfonts}
\usepackage{booktabs} 
\usepackage{threeparttable} 
\usepackage{algorithm}
\usepackage{algorithmic} 
\usepackage{listings}
\usepackage{enumitem}
\usepackage{chngcntr}
\usepackage{lipsum}
\usepackage{subcaption}
\usepackage{authblk}
\usepackage[T1]{fontenc}    
\usepackage{csquotes}       
\usepackage{diagbox}
\usepackage{cite}

\usepackage{setspace}
\usepackage[normalem]{ulem}
\newcommand\erase{\bgroup\markoverwith{\textcolor{red}{\rule[.5ex]{2pt}{0.4pt}}}\ULon}

\usepackage{float}   
\usepackage{caption} 
\makeatletter

\renewcommand{\section}{
  \@startsection{section}
    {1}
    {\z@}
    {10ex plus -1ex minus -.2ex}
    {1.5ex plus.2ex}
    {\Large\normalfont\bfseries}
}
\renewcommand{\subsection}{
  \@startsection{subsection}
    {2}
    {\z@}
    {-3.5ex plus -1ex minus -.2ex}
    {0.2ex plus.2ex}
    {\large\normalfont\bfseries}
}
\makeatother

\usepackage{xurl}

\usepackage{color}

\begin{document}

\noindent{\large\textbf{Emergent intercellular contact-force networks shaped by deformability}}\\

\noindent Koshi Yoshida$^{1,\ast}$, Nen Saito$^{2}$, Shuji Ishihara$^{3,4}$, Kaoru Sugimura$^{4,5,6\ast}$

\begin{enumerate}
 \setlength{\parskip}{0cm} 
 \setlength{\itemsep}{0cm} 
 \setlength{\itemindent}{-4mm}
 \item[$^1$]  Department of Physics, Graduate School of Science, The University of Tokyo, Tokyo 113-0033, Japan
 \item[$^2$]  Life Science Center for Survival Dynamics, Tsukuba Advanced Research Alliance (TARA), University of Tsukuba, Ibaraki 305-8577, Japan
 \item[$^3$]  Department of Basic Science, Graduate School of Arts and Sciences, The University of Tokyo, Tokyo 153-8902, Japan
 \item[$^4$]  Universal Biology Institute, The University of Tokyo, Tokyo 113-0033, Japan
 \item[$^5$]  Department of Biological Sciences, Graduate School of Science, The University of Tokyo, Tokyo 113-0032, Japan
 \item[$^6$]  Department of Computational Biology and Medical Sciences, Graduate School of Frontier Sciences, The University of Tokyo, Chiba 277-8563, Japan 
\end{enumerate}
\noindent $^{\ast}$ Corresponding authors:\\
yoshida-koshi489@g.ecc.u-tokyo.ac.jp (KY), sugimura@bs.s.u-tokyo.ac.jp (KS)

\vspace{10mm}
\noindent\textbf{Abstract}\\
We present a numerical study of contact-force organization in dense assemblies of active, deformable cells. 
Using the Fourier Contour Cell model, 
which enables efficient simulation of large cell deformations, we investigate how cell deformability and active motility shape spatial organization of intercellular forces. 
We find that, in the soft liquid regime at intermediate self-propulsion strength, the contact-force distribution becomes bimodal, reflecting the emergence of a distinct population of weak interactions. 
Persistent-homology analysis further reveals a characteristic force-network structure in which weak tip–tip interactions are enclosed by loops of strong side–side interactions. 
Simulations with elongated but non-deformable cells demonstrate that both the bimodal force distribution and the associated network motif require cell deformability rather than elongated cell shape per se. 
Furthermore, introducing a curvature constraint shows that local cell curvature regulates the strength of weak tip–tip interactions. 
Together, these results demonstrate that cell deformability and local geometric constraints can organize force transmission across dense active cellular assemblies.

\clearpage


\section{Introduction}
The mechanics of multicellular tissues play a crucial role in maintaining tissue structural integrity and regulating biological processes such as embryonic development, wound healing, and cancer progression~\cite{xi_material_2018, trubuil_tissue_2021, ayad_tissue_2019, lim_forced_2024, linke_compressive_2024}.
In these contexts, forces generated by cells and their surrounding extracellular environment are transmitted across cell–cell contacts and regulate diverse cellular functions, including cell motility and cell differentiation~\cite{park_collective_2016, wagh_mechanical_2021, simara_forces_2026}. 

The relationship between force transmission and particle geometry has been extensively studied in inert particulate systems, where particle shape strongly influences the magnitude of contact forces~\cite{zhao_role_2023, radjai_bimodal_1998, tighe_force_2010, behringer_physics_2019}.
For example, in assemblies of elongated particles, side–side contacts preferentially carry strong forces, whereas cap–cap contacts are comparatively weak~\cite{azema_stress-strain_2010, azema_force_2012}.
In living cells, however, this relationship can be bidirectional: forces alter cell shapes, which in turn influence mechanical interactions among cells.
Because studies of particulate systems typically assume rigid particles with fixed shapes~\cite{zhao_role_2023}, how such coupling between deformation and force transmission organizes contact-force networks in assemblies of deformable elements remains poorly understood.

Addressing this question requires a modeling framework that can accommodate large cell deformations while also allowing simulations of sufficiently large systems to resolve the organization of force network.
Existing models face limitations in meeting these requirements.
Vertex-based and self-propelled Voronoi models~\cite{lange_vertex_2025, henkes_dense_2020} represent cells as polygons, which limits the range of cell shapes that can be represented.
Multi-phase-field models~\cite{monfared_multiphase-field_2026, loewe_solid-liquid_2020} can capture highly deformable cell shapes, but their high computational cost limits simulations to relatively small system sizes.
Recently, a computationally efficient model, called a Fourier-Contour Cell model, was proposed~\cite{saito_cell_2024}. 
This model represents cell contours using a Fourier expansion in polar coordinates and enables large-scale simulations while retaining the ability to describe substantial cell deformations.

In addition to a suitable computational model, understanding the organization of intercellular forces requires tools to characterize the spatial structure of force networks.
Topological data analysis has proven useful for extracting mechanical and geometrical properties from heterogeneous assemblies of interacting elements~\cite{wasserman_topological_2018, zheng_topological_2026, tanaka_persistent_2018, hiraoka_hierarchical_2016, minamitani_persistent_2025}.
While it is widely used for analyzing the spatial configuration of disordered particles, it can also characterize topological features of force networks, such as their connectivity and loop structures~\cite{kramar_persistence_2013, kramar_quantifying_2014}. 
Despite this potential, applications of topological data analysis to active matter remain limited and have primarily focused on the spatial organization of particles~\cite{topaz_topological_2015}.

Here, we investigate how cell deformability shapes force transmission in dense assemblies by combining the Fourier-contour cell model with topological data analysis.
We find that, in the soft liquid regime, the distribution of contact forces becomes bimodal, reflecting the coexistence of strong and weak interactions.
Using persistent homology, a method from topological data analysis, we show that strong force loops enclosing weaker interactions emerge.
Furthermore, we demonstrate that this bimodality and the associated topological structures of the force network arise from cell deformability rather than cell elongation, and are governed by geometric constraints on cell shape, in particular limitations on local curvature of the cell contour.
Together, these results reveal a deformability-driven mechanism for organizing force networks in dense active cellular systems.


\section{Materials and Methods}

\setlength{\abovedisplayskip}{10pt} 
\setlength{\belowdisplayskip}{10pt}

\subsection{Model description}~\label{Methods:Model}
We consider a system of $N$ deformable cells that interact with each other through excluded volume interaction. 
Following our previous study~\cite{saito_cell_2024, pal_dense_2026}, the cell contour of the $i$-th cell is expressed in a polar coordinate, where the cell center $\vb{r}_c^i$ of $i$-th cell is set as the origin and the radius in $\theta$ direction is expressed as $R^i(\theta)$, which is further expanded as a finite Fourier series up to the $M$-th mode as,
\begin{align}
    R^i(\theta) = R_0 \sum_{n = 0}^M \left[ a_n^i \cos{n(\theta - \theta^i)} + b_n^i \sin{n(\theta - \theta^i)} \right],
\end{align}
where $\theta_i$ denotes the migrating direction of $i$-th cell.
To guarantee the uniqueness of the origin of the expansion, we set as $a_1^i = b_1^i  = 0$. 
Note that $b_0^i$ can be set as $b_0^i=0$ since $b_0^i\sin 0 =0$. 
Furthermore, a field function $\phi^i(\bf r)$ is introduced as $\phi^i({\bf r})=1/2+(1/2)\tanh \left( \frac{R^i(\hat{\theta})-|{\bf r -r_c^i}|}{\epsilon} \right)$, where $\epsilon$ is a small parameter to regulate the interface width and $\hat{\theta} = \mbox{arg} ({\bf r}-{\bf r_c^i})$. 
The $i$-th cell interior $\Omega_i$ is then defined as the region that satisfies $\phi^i({\bf r}) > 1/2$ .

The total Hamiltonian of the system is given as
$\textit{H} = \textit{H}_\text{int} + \textit{H}_\text{l}$. 
The first term $\textit{H}_\text{int} = \sum_{i < j} \int \phi^i \phi^j \dd {\bf r}$ represents the potential energy for the excluded volume effect penalizing overlapping area between $i$-th and $j$-th cells, 
and the second term, $\textit{H}_{l} = \eta\sum_i L_i$, denotes the surface-energy term associated with the cortical tension $\eta$, where $L_i$ is the interface length of cell $i$.

The time evolution is given by the following equations.
\begin{align}
&\dot{\vb{r}}_c^i = -\mu_r \frac{\partial \mathcal{H}}{\partial \vb{r}_c} + v_0 \vb{e}_{\theta^i} \label{eq:rc}\\
&\dot{\theta}^i = -\mu_\theta \frac{\partial \mathcal{H}}{\partial \theta^i} + \sqrt{2D_\theta}\xi^i \label{eq:torque}\\
&\dot{a}_n^i = -\mu_{ab}\left(\frac{\partial \mathcal{H}}{\partial a_n^i}  - C\frac{a_n}{4a_0^2 + \sum_{m=2}^M(a_m^{i 2} + b_m^{i 2})}\right) \qquad (n = 2, \dots, M) \label{eq:an}\\
&\dot{b}_n^i = -\mu_{ab}\left(\frac{\partial \mathcal{H}}{\partial b_n^i}  - C\frac{b_n}{4a_0^2 + \sum_{m=2}^M(a_m^{i 2} + b_m^{i 2})}\right) \qquad (n = 2, \dots, M) \label{eq:bn}
\end{align}
where $C = 2a_0 \frac{\partial \mathcal{H}}{\partial a_0}
+ \sum_{n=2}^M \left( a_n \frac{\partial \mathcal{H}}{\partial a_n}
+ b_n \frac{\partial \mathcal{H}}{\partial b_n} \right)$ and $\xi^i$ denotes normalized Gaussian white noise. 
The second term $v_0\vb{e}_{\theta^i}$ in 
eq.~\eqref{eq:rc} represents self-propulsion of the particle in the direction 
$\vb{e}_{\theta^i}=(\cos\theta^i,\sin\theta^i)$ 
at a prescribed speed $v_0$.
Taking the limit $\epsilon \to 0$, the derivatives
$\frac{\partial \mathcal{H}}{\partial \vb{r}_c}, \frac{\partial \mathcal{H}}{\partial \theta^i}, \frac{\partial \mathcal{H}}{\partial a_n^i}$, and $\frac{\partial \mathcal{H}}{\partial b_n^i}$ are further calculated as:
\begin{eqnarray}
\frac{\partial \mathcal{H}}{\partial \vb{r}_c} &=& -\sum_{j\neq i} \int_{-\pi}^{\pi}d\theta \phi^j (R'^{i}\vb{e}_\perp-R^i\vb{e}_\theta) \label{eq:dHdrc}\\
\frac{\partial \mathcal{H}}{\partial \theta^i}&=& -\sum_{j\neq i} \int_{-\pi}^{\pi}d\theta \phi^j R'^{i}R^i\\
\frac{\partial \mathcal{H}}{\partial a_n^i}&=&\eta R_0 \int_{-\pi}^{\pi}d\theta \frac{R^i \cos n(\theta -\theta^i)-R'^i n \sin n(\theta -\theta^i)}{\sqrt{(R^i)^2+(R'^i)^2}} \nonumber\\
&+& \sum_{j\neq i} \int_{-\pi}^{\pi}d\theta \phi^j R^i R_0 \cos n(\theta -\theta^i)\\
\frac{\partial \mathcal{H}}{\partial b_n^i}&=&\eta R_0 \int_{-\pi}^{\pi}d\theta \frac{R^i \sin n(\theta -\theta^i)+R'^i n \cos n(\theta -\theta^i)}{\sqrt{(R^i)^2+(R'^i)^2}} \nonumber\\
&+& \sum_{j\neq i} \int_{-\pi}^{\pi}d\theta \phi^j R^i R_0 \sin n(\theta -\theta^i), \label{eq:dHdbn}
\end{eqnarray}
where 
$\vb{e}_{\theta} = (\cos\theta, \sin\theta)$ and $\vb{e}_\perp =(-\sin\theta, \cos\theta)$. 
In eqs.~\eqref{eq:an} and \eqref{eq:bn}, the second terms are introduced to maintain the conservation of area to $\pi R_0^2$, namely:
\begin{equation}
\label{eq:area_cons}
    a_0^2 + \sum_{n=2}^{M} \frac{a_n^2 + b_n^2}{2} = 1.
\end{equation}
After obtaining $\{ a_n^i\}$ and $\{ b_n^i\}$ for $n=2,\dots,M$ at each time step, $a_0^i$ is updated as  $a_0^i=\sqrt{1 - \sum_{n = 2} \frac{(a_n^i)^2 + (b_n^i)^2}{2}}$.
Note that $a_0^i$ can be interpreted as an independent variable obeying eq.~\eqref{eq:an}, however we impose this update to avoid the accumulation of numerical errors in area conservation instead of considering eq.~\eqref{eq:an} for $a_0^i$.
In addition, to prevent self-intersections of the cell contour, the constraint $\sum_{n=2}^{M} \sqrt{(a_n^i)^2 + (b_n^i)^2} < \sqrt{\frac{2}{3}}$ is imposed.

The parameters are set as $R_0 = 1, M = 6, \mu_r = 1, \mu_\theta = 1, \mu_{ab} = 0.1$, $D_\theta=0.01$ unless otherwise specified.

\subsection{Numerical calculation}~\label{Methods:NumericalCalculation}
The time evolution equations \eqref{eq:rc} - \eqref{eq:bn} are numerically solved using the Euler-Maruyama method. 
For the integral with respect to $\theta$ in eqs.~\eqref{eq:dHdrc}--\eqref{eq:dHdbn}, the intersection points of two interacting-cell contours need to be identified to calculate the interactions among cells. 
In our previous study~\cite{saito_cell_2024}, these intersection points were searched based on whether the point on the $i$-th cell contour, $\vb{r}_c^i + R^i(nd\theta)\vb {e}_{nd\theta}$, was inside or outside $j$-th cell for a given $d\theta$ and $n = 0, 1, 2, \cdots$. 
However, this scheme cannot estimate the intersections precisely. 
Consequently, it yields spurious large fluctuation in the dense condition even without the self-propulsion and noises, while it can be reduced or eliminated using smaller $\dd \theta$. 
In the present study, the intersections of two interacting-cell contours are precisely identified by using a binary search within pre-defined error of angle.  
A detailed explanation of the algorithm is provided in Supplementary Information{\color{blue}{~S1}}.

In the simulations, the system consists of $N = 256$ cells in a periodic box of $L_x:L_y = 2:\sqrt{3}$ with the packing fraction fixed at $\pi R_0^2 N/L_x L_y = 1$. 
The Euler--Maruyama method is used with a time step of $dt = 0.001$.
For the numerical integration with respect to $\theta$, the interval $[0, 2\pi)$ is discretized using 80 uniformly spaced points together with the intersection points identified by the binary search algorithm described above. 
Each subinterval $\dd\theta$ 
between two consecutive points defines one integration bin.
As the initial condition, cells are arranged on a hexagonal lattice, and their orientation angles are set randomly.

\subsection{Orientational order parameter}~\label{Methods:OrderParameter}
The local orientational order parameter for cell $j$ is given by
$|\Psi_{6,j}|= \left| \frac{1}{N_{\mathrm{nn},j}} \sum_{k\in\mathrm{nn}(j)} \exp(i6\theta_{jk}) \right|$,
where $N_{\mathrm{nn},j}$ is the number of nearest neighbors of cell $j$, as determined by Delaunay triangulation, and $\theta_{jk}$ is the angle of the relative position vector between cells $j$ and $k$ with respect to a fixed reference axis. The global orientational order parameter $|\Psi_6|$ is defined as $|\Psi_6| = |\frac{1}{N}\sum_j \Psi_{6, j}|$~\cite{loewe_solid-liquid_2020}.

\subsection{Intercellular force and its visualization}~\label{Methods:Force}
The force resulting from the exclusion volume interactions exerted from the $j$-th cell to the $i$-th cell is given by $\vb{F}^{ij} = \mu_r \int \dd \vb{r} \phi^j \nabla\phi^i(\vb{r})$, which corresponds to the right hand side of eq.~\eqref{eq:dHdrc} except for the factor $\mu_r$. 
In the actual implementation, we replace $\vb{F}^{ij}$ with $\frac{1}{2}\left(\vb{F}^{ij} - \vb{F}^{ji}\right)$ to precisely satisfy the action-reaction principle.
To visualize the force network, edges are drawn between the centroids of contacting cell pairs and colored according to the magnitude of the contact force, although the actual force is directed perpendicular to the interface, rather than the center-to-center direction.

An advantage of the Fourier-contour cell model is that it provides a direct correspondence between the magnitude of contact forces and the geometrical overlap between neighboring cells (Supplementary Information{\color{blue}{~S2}}).
This enables intercellular forces to be interpreted in terms of the resulting contact geometry, even in the presence of active fluctuations.

\subsection{Persistent homology}~\label{Methods:PersistentHomology}
The persistent diagram (PD) of the force network is constructed by employing the inverse Vietoris-Rips filtration~\cite{aktas_persistence_2019}. 
Consider a weighted graph $G=(V,E)$ with a nonnegative weight function $F:E\to\mathbb{R}$, where $F(e)$ represents the force magnitude between the contacting cells connected by edge $e$.
$1$-skelton is defined as $G_\delta=(V_\delta,E_\delta )$, where $V_\delta = V$ and $ E_\delta = \{e | F(e) \geq \delta\}$. 
Here, $Cl(G_\delta)$ is a clique complex made from $G_\delta$.
Varying the threshold $\delta$ from $F_{\max}$ to $0$, where $F_{\max}=\max_{e\in E}F(e)$, yields the following filtration:
\begin{align}
    \operatorname{Cl}(G_{F_{\max}})
    \subseteq \cdots \subseteq
    \operatorname{Cl}(G_\delta)
    \subseteq \cdots \subseteq
    \operatorname{Cl}(G_0) = 
    \operatorname{Cl}(G),
    \qquad
    F_{\max}\geq\delta\geq0.
\end{align}
Here, \(\operatorname{Cl}(G_\delta)\) denotes the clique complex constructed from $G_\delta$.

The $H_1$ persistence diagram is obtained by tracking loops throughout the filtration and plotting their birth–death pairs (Figure {\color{blue}{S7}}). 
Birth denotes the threshold at which a loop first encloses a hole, whereas death denotes the threshold at which added simplices fill the enclosed hole.


\section{Results}
\subsection{Dynamics and phase diagram of the system}~\label{Results:PhaseDiagram} 
We first characterized the spatial organization and temporal dynamics of the system to obtain an overall picture of the dense active assembly of deformable cells.
Using the Fourier-contour cell model~\cite{saito_cell_2024}, the shape of each cell is represented as a single-valued function describing the radial distance to the contour, expressed as a Fourier series up to the $M = 6$-th mode (see Sec.~\ref{Methods:Model} for details). 
Cell deformation is penalized by surface energy, with lower cortical tension corresponding to greater deformability.
Simulations were performed over a range of self-propulsion strengths $v_0$ and cortical tensions $\eta$, which serve as the control parameters in this study.

\begin{figure}[H]
    \centering
    \includegraphics[width=\linewidth]{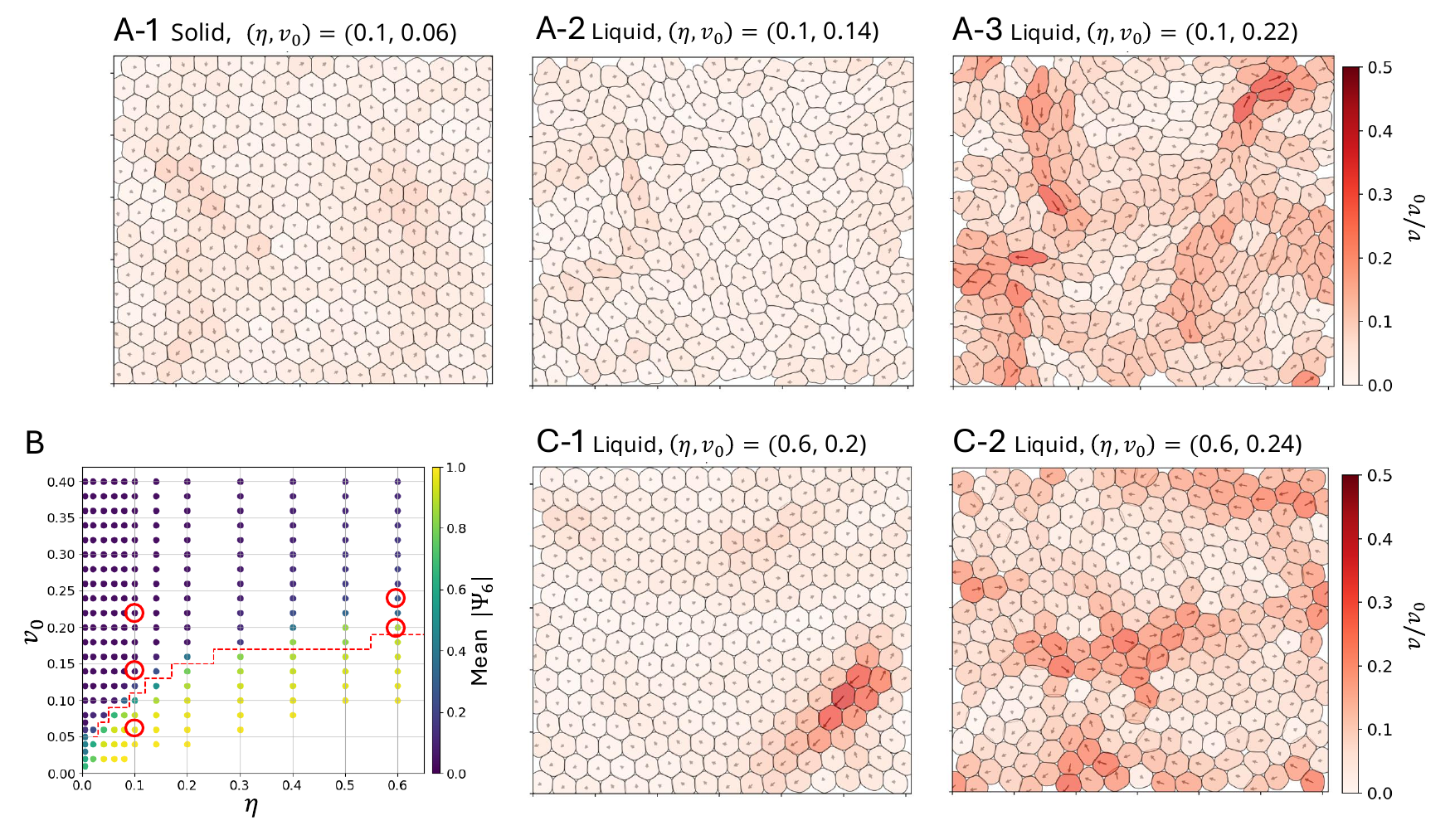}
    \vspace{-10pt}
    \caption{Dynamics and phases at high packing density for different cortical tension and self-propulsion velocity.
    (A-1)–(A-3) Typical snapshots from numerical simulations with weak cortical tension. Cells are colored according to their normalized velocity (color-bar). The system exhibits a solid phase in (A-1) and a liquid phase in (A-2) and (A-3). The parameters used (cortical tension $\eta$ and self-propulsion velocity $v_0$) are indicated at the top of each panel.
    (B) Phase diagram in the $(\eta, v_0)$ parameter space. The global orientational order parameter $|\Psi_6|$ (Sec.~\ref{Methods:OrderParameter}) is shown as an indicator of phase (color-bar). 
    The phase boundary is depicted by the red dashed line.
    The red circles mark the locations in parameter space corresponding to the snapshots shown in (A) and (C).
    (C-1)–(C-2) Typical snapshots from numerical simulations with strong cortical tension in the liquid regime close to the solid–liquid phase boundary. Cells are colored according to their normalized velocity as in (A).
    }
    \label{fig:Phase}
\end{figure}

Figure~\ref{fig:Phase}A-1--A-3 show typical snapshots at low cortical tension ($\eta = 0.1$) for $v_0 = 0.06$, $0.14$, and $0.22$. 
Cells are colored according to their instantaneous speed, computed from their displacement over $\Delta t = 10$.
For small propulsion strength $v_0 = 0.06$, cells remain near their initial positions on a hexagonal lattice (Figure~\ref{fig:Phase}A-1).
When $v_0$ exceeds a threshold value, a liquid phase emerges in which cells move more extensively by rearranging their positions relative to their neighbors (Figure~\ref{fig:Phase}A-2 and A-3).
The solid–liquid phase boundary is indicated by a red dashed line in Figure~\ref{fig:Phase}B. 
The solid phase is determined by the effective diffusion coefficient $D_\text{eff} = D_s/D_0 < 0.5 \times 10^{-4}$,
where $D_s$ is the diffusion coefficient measured by mean-squared displacement (MSD) and $D_0 = v_0^2/2D_\theta$ is the reference diffusion coefficient of single cell (Figure{\color{blue}{~S1}}). 
In the liquid phase, cells are highly deformed (Figure~\ref{fig:Phase}A-2 and A-3; Movie{\color{blue}{~1}}), and the initially ordered arrangement rapidly becomes disordered, as indicated by the rapid decrease in the global orientational order parameter $|\Psi_6|$ (Figure~\ref{fig:Phase}B; Figure {\color{blue}{S2}}A, B, E).

The MSD also reveals a prolonged super-diffusive regime in the liquid phase (Figure{\color{blue}{~S1}}), which persists longer than $D_\theta^{-1}$ expected for active Brownian particles without a torque term~\cite{mandal_extreme_2020, szamel_extremely_2024}. 
This prolonged super-diffusive regime can be attributed to interactions with neighboring cells, which effectively generate a restoring torque in the equation of motion (eq.~\eqref{eq:torque}).

Near the solid–liquid phase boundary in the rigid regime, the system exhibits intermittent dynamics characterized by bursts of motion separated by quiescent periods (Figure~\ref{fig:Phase}C, Figure{\color{blue}{~S2}}C, D, G, Figure {\color{blue}{S3}}; Movie{\color{blue}{~2}}; Supplementary Information{\color{blue}{~S3}}).  
Consistent with previous observations in a multi-phase-field model~\cite{loewe_solid-liquid_2020}, system-scale intermittency was suppressed with increasing system size (Figure~{\color{blue}{S2}}F). 
In contrast, intermittent dynamics at the single-cell scale persisted even in large systems (Figure~{\color{blue}{S3}}E, F).
In the following, we focus primarily on the soft liquid regime, outside the intermittent region near the solid–liquid phase boundary.

\subsection{Deformability-dependent bimodal distribution of contact forces}~\label{Results:ForceDistribution} 
Having characterized the phases and dynamical behaviors of the dense system of active deformable cells, 
we next investigated how deformability and persistent 
self-propulsion shape mechanical interactions between cells.
The contact force between a pair of contacting cells was computed as described in Sec.~\ref{Methods:Force}.
Importantly, in our model, the magnitude of the contact force between cells $i$ and $j$ directly reflects the geometry of their contact: $|\vb{F}^{ij}| = \mu_r|\int \dd s \vb{n}^i(r)|$,
where the integral is taken over the overlap region between the two cells; 
the force magnitude is proportional to the distance between the two
intersection points of their contours (Supplementary Information{\color{blue}{~S2}}).
The characteristic time from contact formation to detachment is significantly longer than the time scale used to compute the intercellular force (see Figure {\color{blue}{S4}} for the susceptibility and bond-breaking function).
This separation of timescales allows the contact-force distribution to be defined robustly despite stochastic active fluctuations.

\begin{figure}[H]
    \centering
    \includegraphics[width=0.8\linewidth]{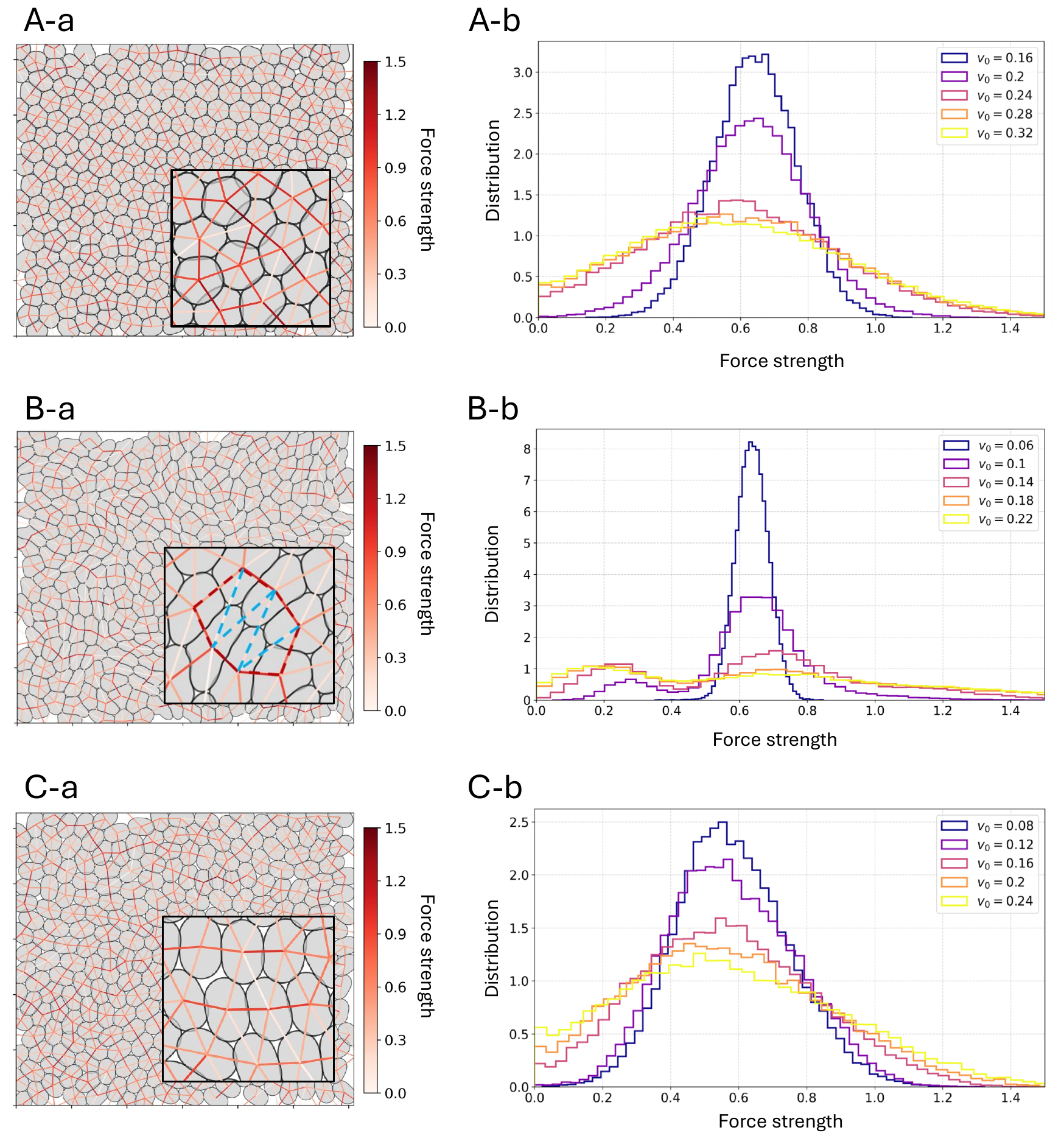}
    \caption{Force distributions under rigid, soft, and non-deformable conditions. 
    (A-a), (B-a), (C-a) Snapshots from the simulations showing the intercellular force network. Edges are drawn between the centroids of contacting cell pairs and colored according to the magnitude of the contact force (color bar); the edges indicate contact pairs rather than force directions, as the contact forces act locally normal to the cell contours. The parameters used are $(\eta, v_0) = (0.6, 0.24)$ (A-a; rigid), $(\eta, v_0) = (0.1, 0.14)$ (B-a; soft), and $(a_2, v_0) = (0.2, 0.16)$ (C-a; non-deformable).
    In (B-a), strong interactions forming loops and the weak interactions enclosed by these loops are highlighted by red and light-blue dashed lines, respectively.
    (A-b), (B-b), (C-b) Corresponding force distributions. Different colors indicate different values of the self-propulsion strength $v_0$ used in the simulations. 
    }
    \label{fig:fig2}
\end{figure}

Our analysis revealed that the force distribution strongly depends on cortical tension 
and self-propulsion strength.
For rigid cells with high cortical tension ($\eta = 0.6$; Figure~\ref{fig:fig2}A-a), the distribution at low $v_0$ peaks around $0.65$ with nearly symmetric tails (blue line in Figure~\ref{fig:fig2}A-b).
As $v_0$ increases, the distribution broadens and the peak shifts toward smaller forces.
The tails decay faster than exponentially (see Figure {\color{blue}{S5}} for the semi-log plot),
similar to the force distribution of frictionless disks on a triangular lattice, where the contact force distribution is peaked at a finite value and exhibits super-exponential decay~\cite{tighe_force_2010, snoeijer_force_2004}. 
A qualitatively different force distribution is observed for soft cells ($\eta = 0.1$).
At small $v_0$, the force distribution is unimodal and centered around $0.65$, as in the rigid case (blue line in Figure~\ref{fig:fig2}B-b). 
However, in contrast to the rigid case, increasing $v_0$ leads to the emergence of a second peak around $0.2$, resulting in a bimodal force distribution at intermediate self-propulsion strengths (purple and magenta lines in Figure~\ref{fig:fig2}B-b).
This bimodality is observed only at packing fractions above $0.95$, irrespective of the protocol used to elevate the cell density. 
With further increase in $v_0$, the force distribution broadens over a wide range and the bimodality becomes less pronounced (orange and yellow lines in Figure~\ref{fig:fig2}B-b).

In the soft regime, cells become strongly deformed (Figure~\ref{fig:fig2}B-a), raising the question of whether the observed weak interactions arise from the cell shape itself or from the ability of cells to deform.
To distinguish between these possibilities, we performed simulations with elongated but non-deformable cells (Figure~\ref{fig:fig2}C). 
Specifically, the second Fourier mode was set to $a_2 = 0.2$, and all Fourier modes were subsequently fixed at their initial values throughout the simulation. 
The resulting force distribution is unimodal and broadens as the self-propulsion strength $v_0$ increases (Figure~\ref{fig:fig2}C-b), with modest variations in $a_2$ producing qualitatively similar results.
These results demonstrate that cell elongation alone is insufficient to generate the bimodal force distribution; instead, cell deformability is essential.

The bimodal force distribution remains largely unchanged across different values of $D_\theta$, including the limit $D_\theta = 0$ (Figure {\color{blue}{S6}}). 
The bimodality is also preserved in the absence of the mechanical torque term.
The only exception is when $D_\theta$ becomes large enough that the persistence length $l_p = v_0/D_\theta$ is comparable to the particle size (Figure {\color{blue}{S6}}, $l_p = 10^0$), where the system is no longer fluidized.
These results indicate that the bimodal force distribution is a robust consequence of cell deformability in the fluidized regimes and does not depend sensitively on orientational noise or mechanical torque.

Collectively, our data demonstrate that cell deformability, persistent self-propulsion, and high packing fraction are all essential for shaping the mechanical interactions in this active dense system.

\subsection{Geometric Analysis of the Force Network Using Persistent Homology}
~\label{Results:TDA} 
Although the force distribution reveals the coexistence of strong and weak interactions, it does not specify how these interactions are geometrically organized within the force network. 
To uncover the geometric structure underlying the force network, we employed persistent homology, a method from topological data analysis that has been applied to granular materials~\cite{tanaka_persistent_2018, kramar_persistence_2013}.
Specifically, a persistence diagram (PD) is constructed using an inverse Vietoris–Rips filtration (Sec.~\ref{Methods:PersistentHomology}; Figure {\color{blue}{S7}}A)~\cite{aktas_persistence_2019}. 
Starting from the strongest interaction, edges are sequentially added in descending order of contact-force magnitude, generating a series of subnetworks at different thresholds. 
The birth and death of a loop are defined as the thresholds at which the loop appears and becomes filled, respectively. 
Each loop is therefore represented as a point on the PD with coordinates corresponding to its birth and death thresholds~(Figure {\color{blue}{S7}}B). 
Loops with a larger difference between their birth and death thresholds appear farther from the diagonal in the PD, reflecting a greater separation between the force scales associated with loop formation and filling.
In this way, the PD encodes geometric information about the force network, including the spatial arrangement of strong and weak interactions.

\begin{figure}[H]
    \centering
    \includegraphics[width=\linewidth]{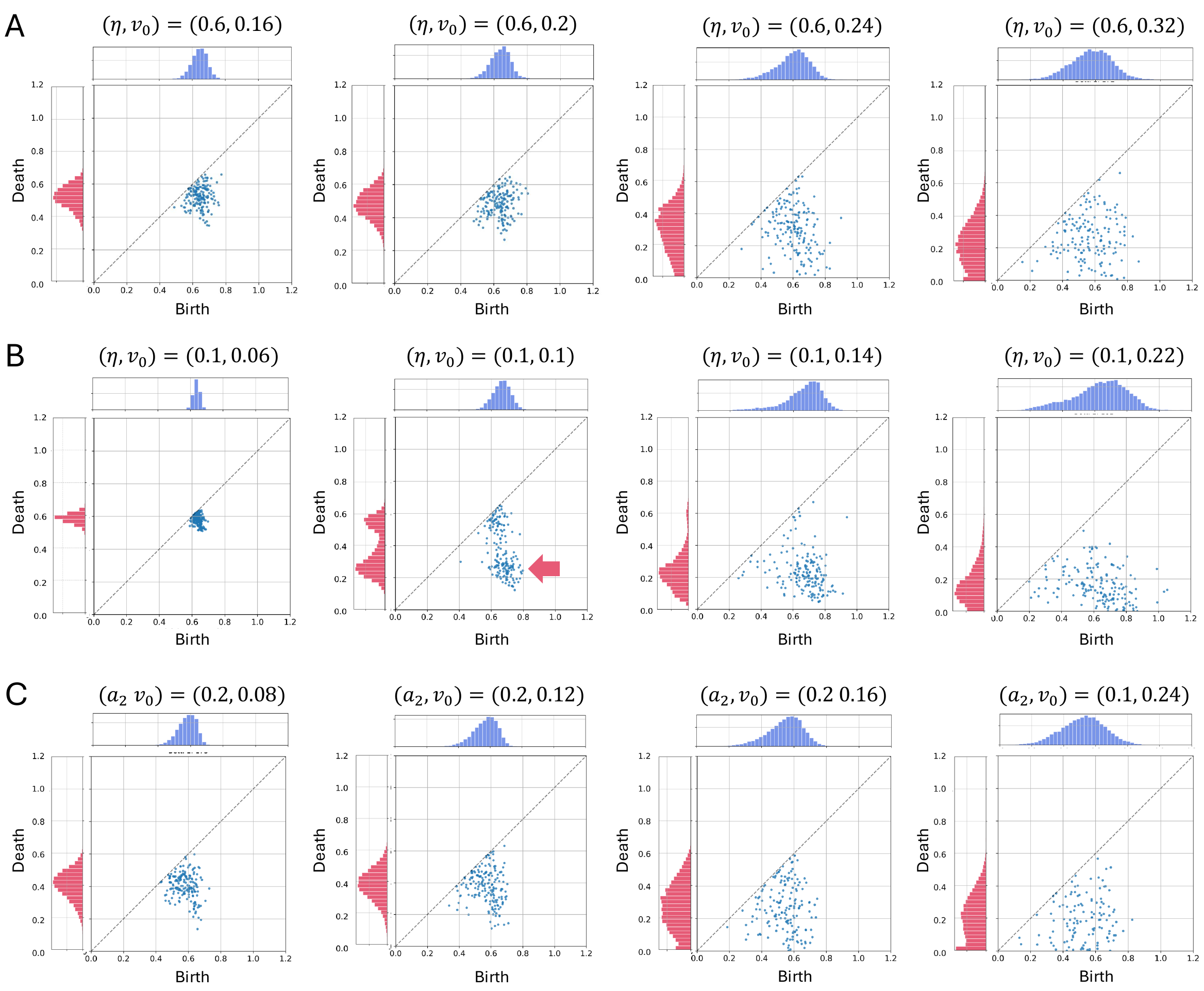}
    \caption{Persistence diagrams (PDs) under rigid, soft, and non-deformable conditions. 
    (A–C) $\eta = 0.6$ (A; rigid), $\eta = 0.1$ (B; soft), and $a_2 = 0.2$ (C; non-deformable). 
    Each column corresponds to different values of the self-propulsion strength $v_0$. 
    Contact force strengths averaged within $\Delta t = 1$ are used for PDs.
    Frequencies of birth and death are shown above and to the left of the PDs, respectively. 
    These distributions are calculated from PDs obtained at 100 different time points.  
    (A, B) At small $v_0$, points are clustered near the diagonal (leftmost panels).
    At intermediate $v_0$, the arrow in the second panel of (B) indicates the distinct off-diagonal cluster.
    As $v_0$ increases, the PDs for both rigid and soft cells become increasingly dispersed (right two panels), consistent with the broadening of the contact-force distributions (Figure~\ref{fig:fig2}A-b, B-b).
    (C) The PDs of non-deformable cells show patterns similar to those of rigid cells. 
    }
    \label{fig:PD}
\end{figure}

Figures~\ref{fig:PD}A and B show the PDs for force networks at cortical tensions $\eta = 0.6$ and $0.1$, respectively, for varying self-propulsion strengths $v_0$. 
The birth and death distributions, shown above and to the left of each PD, respectively, are obtained by pooling data from 100 time points for each condition.
In the solid phase where $v_0$ is small, both rigid and soft cells exhibit a single cluster located near the diagonal line (leftmost panels in Figure~\ref{fig:PD}A, B).
As $v_0$ increases and the system enters the liquid phase, the PDs reveal distinct network structures for rigid and soft cells.
For rigid cells, the cluster gradually dissolves without leaving a distinct structural signature (Figure~\ref{fig:PD}A).
In sharp contrast, for soft cells, an off-diagonal cluster centered around $(\text{Birth}, \text{Death}) = (0.7, 0.2)$ emerges, most evidently at intermediate self-propulsion strength (indicated by an arrow in the second panel of Figure~\ref{fig:PD}B).
Importantly, this off-diagonal cluster is strongly suppressed in the absence of cell deformability (Figure~\ref{fig:PD}C; the secondary peak in the death distribution observed for soft cells at intermediate $v_0$ is absent).

In the FCC model, this off-diagonal cluster geometrically corresponds to a configuration in which strong side–side contacts surround weak tip–tip contacts (Figure~\ref{fig:fig2}B-a).
Furthermore, the accumulation of points in a distinct off-diagonal region of the PD suggests that this characteristic loop motif recurrently forms in assemblies of soft cells.

\begin{figure}[H]
    \centering
    \includegraphics[width=0.5\linewidth]{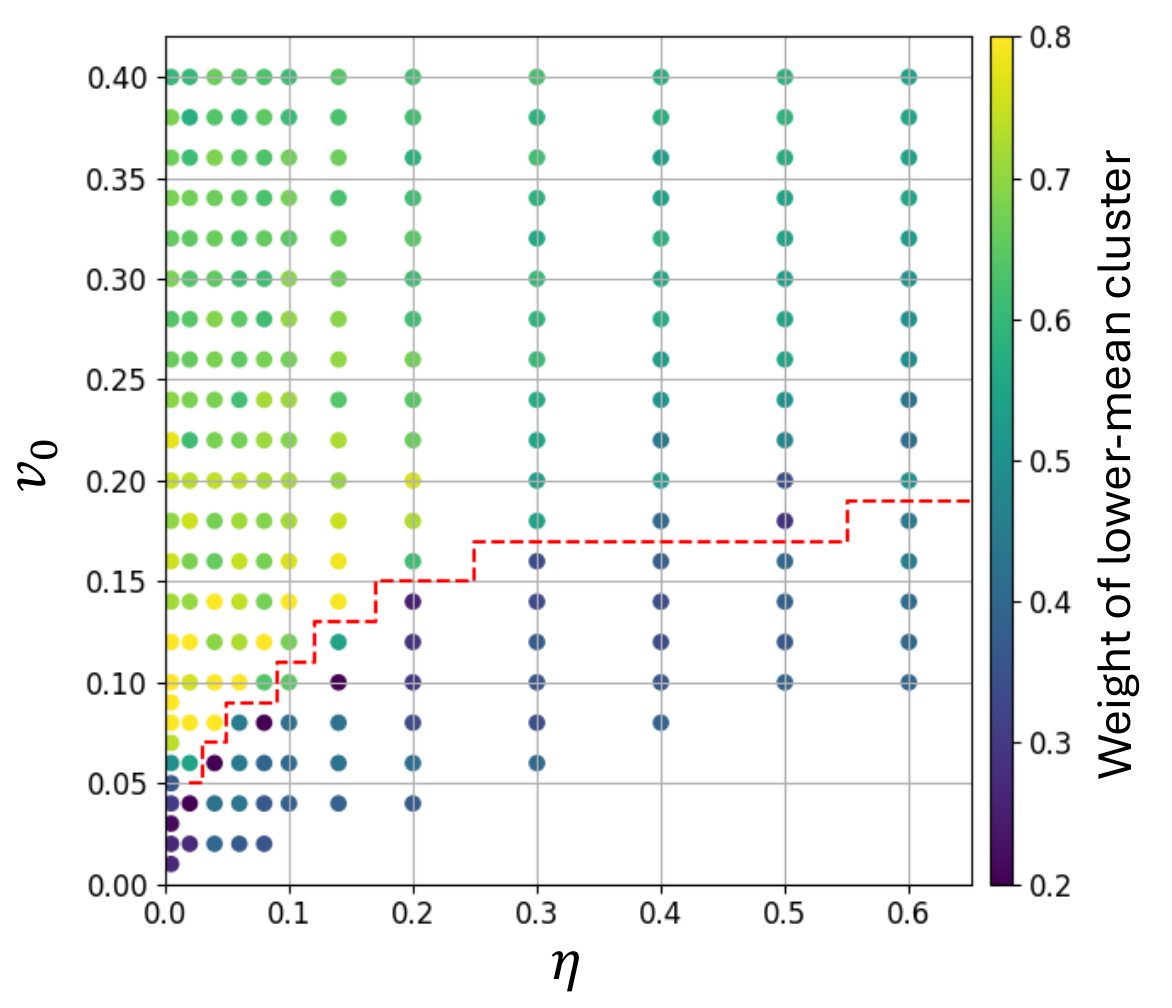}
    \caption{Weight of the lower-mean cluster in the distribution of death thresholds in the persistence diagram (PD). A Gaussian mixture model is applied to the distribution of death thresholds to identify the lower-mean cluster associated with the off-diagonal cluster in the PD.
    }
    \label{fig:fig4}
\end{figure}

To systematically compare the parameter dependence of the PDs, specifically, the formation of the off-diagonal cluster, with the phase diagram, we applied a Gaussian mixture model to the distribution of death thresholds, with the number of clusters fixed to two, and plotted the weight of the lower-mean cluster on the phase diagram.
This analysis shows that the weight of the lower-mean cluster is elevated under soft, liquid conditions, particularly just above the solid–liquid phase boundary (Figure~\ref{fig:fig4}), reflecting the accumulation of points in the off-diagonal region of the PD (the second panel of Figure~\ref{fig:PD}B).
Overall, the topological data analysis reveals that, in the soft liquid regime, the force network develops a distinct topological structure that correlates with the bimodality of the force distribution.

\subsection{Cell contour curvature underlies weak interactions among soft cells}
~\label{Results:Curvature} 
Finally, we investigated the mechanism by which weak contact forces, which are associated with the off-diagonal cluster in the PD, emerge and give rise to the bimodal force distribution.
Since cell deformability is essential for this phenomenon, 
we vary the number of Fourier modes, \(M\), which determines the resolution of the cell-contour representation, and the constraint on cell contour curvature, which controls cell deformability.

\begin{figure}[H]
    \centering
    \includegraphics[width=\linewidth]{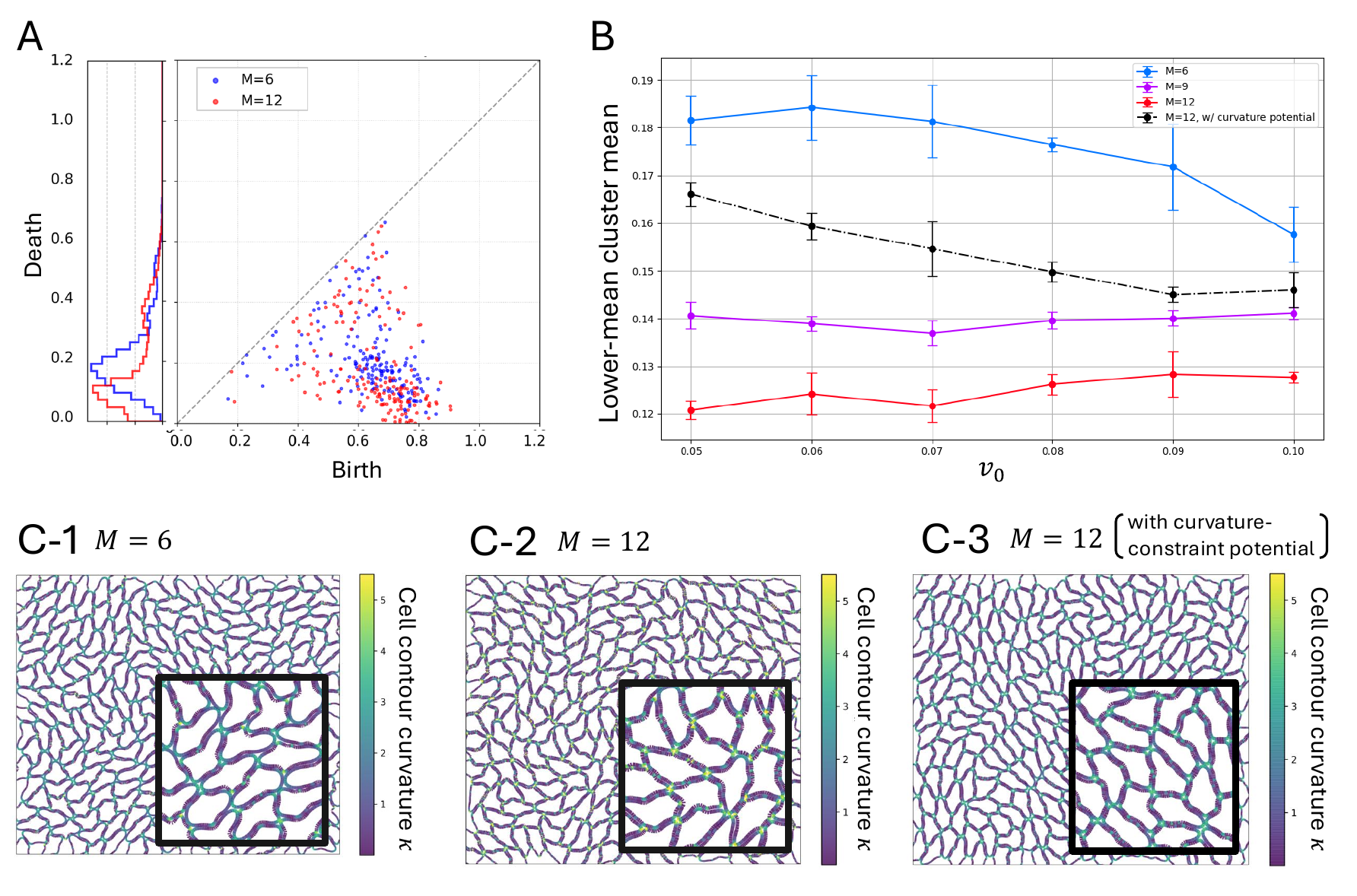}
    \caption{Effect of Fourier modes and the curvature-constraint potential on the force network and cell-contour geometry.
    (A) Persistence diagram (PD) at $(\eta, v_0) = (0.005, 0.09)$ for different numbers of Fourier modes, with the frequency of death threshold values shown on the left ($M = 6$, blue; $M = 12$, red).
    Contact force strengths averaged within $\Delta t = 1$ are used to construct the PDs. 
    (B) Dependence of the mean death threshold value of the lower-mean cluster on the number of Fourier modes and the curvature-constraint potential, averaged over three independent simulation runs for each condition. 
    (C) Snapshots from simulations at $(\eta, v_0) = (0.005, 0.09)$, with cell-contour curvature $\kappa$ shown using a color map: $M = 6$ (C-1); $M = 12$ (C-2); and $M = 12$ with the curvature-constraint potential ($\gamma = 3$, $\kappa_0 = 4.2$) (C-3).
    }
    \label{fig:fig5}
\end{figure}

We first investigated whether and how increasing $M$ alters the PD and the force distribution.
Simulations in this section were performed under extremely weak cortical tension ($\eta = 0.5 \times 10^{-2}$), which facilitates the formation of the off-diagonal cluster in the PD.
Figure~\ref{fig:fig5}A shows overlaid PDs obtained at the same $(\eta, v_0)$ for $M = 6$ (blue) and $M = 12$ (red). 
Increasing $M$ shifts the off-diagonal cluster downward and slightly to the right.
This downward shift on the PD indicates that the weak interactions enclosed by strongly interacting cell pairs become even weaker as $M$ increases (Figure~\ref{fig:fig5}B), which is also reflected in corresponding changes in the force distribution (Figure {\color{blue}{S8}}).

This behavior can be understood from the model formulation, in which the maximum curvature of the cell contour is controlled by the number of Fourier modes.
When the cell contour is colored according to its local curvature, regions of high curvature become increasingly localized at sharp tips as $M$ increases, with the tip curvature being around $4.0$ for $M = 6$ and exceeding $5.0$ for $M = 12$ (Figure~\ref{fig:fig5}C-1, C-2).
As a result, tip–tip contacts become sharper and geometrically more localized, thereby reducing the overlap between neighboring cells and weakening the corresponding contact forces.
These observations suggest that local cell-contour curvature may contribute to setting the strength of weak tip–tip interactions.

To test the role of local curvature, we introduced an additional potential that constrains cell-contour curvature above a threshold. 
This potential was motivated by the fact that the maximum curvature of the cell contour in biological cells is constrained by the elasticity of the plasma membrane and the underlying cortex~\cite{peter_bar_2004, carman_bar_2018}.
We considered the curvature-constraint potential as follows:
\begin{align}
    H_\text{curv}^i &= \gamma \oint \dd s \psi_{\kappa_0}(\kappa) (\kappa - \kappa_0)^2 
\end{align}
where $\psi_{\kappa_0}(\kappa)$ is an indicator function that equals $0$ for $\kappa<\kappa_0$ and $1$ for $\kappa\geq\kappa_0$.
Here, $\kappa$ denotes the curvature along the cell contour, and $\kappa_0$ is the threshold above which curvature is penalized. 
Unlike the Helfrich potential widely used to model cell membranes~\cite{winterhalter_bending_1992}, this potential acts only when the local curvature exceeds the threshold $\kappa_0$.
The threshold $\kappa_0$ was set to $4.2$ based on the distribution of $\kappa$ (Figure {\color{blue}{S9}}; Supplementary Information{\color{blue}{~S4}}).
Modest changes in $\kappa_0$ did not qualitatively affect the results.
The value of $\gamma$ was chosen to maintain numerical stability.

Introducing the curvature-constraint potential to numerical simulations with $M = 12$ restricts the tip curvature to $\kappa < 4.2$ (Figure~\ref{fig:fig5}C-3), comparable to that in the $M = 6$ case without this potential (Figure~\ref{fig:fig5}C-1), thereby limiting the formation of extremely sharp tips characteristic of the unconstrained $M = 12$ case.
Under these conditions, the mean death threshold of the off-diagonal cluster in the PD shifts toward that observed at smaller $M$ (Figure~\ref{fig:fig5}B).
The incomplete recovery of the death threshold may reflect residual differences in cell geometry between the $M = 6$ case and the $M = 12$ case with the curvature-constraint potential, particularly in the smoothness of contact surfaces between neighboring cells.
Consistent with the shift on the PD, the smaller-force peak in the bimodal force distribution also shifts toward the position observed for $M = 6$ (Figure {\color{blue}{S8}}).
These results support that local cell-contour curvature contributes to setting the strength of weak tip–tip interactions and, consequently, the position of the smaller-force peak in the bimodal force distribution.

\section{Discussion}
Force transmission in dense particulate systems has been extensively studied in granular materials, where non-deformable particles passively respond to external loading and develop heterogeneous force networks~\cite{zhao_role_2023, radjai_bimodal_1998, tighe_force_2010, behringer_physics_2019, azema_stress-strain_2010, azema_force_2012}. 
In contrast, cells are deformable and self-propelled, allowing cell shape and intercellular mechanical interactions to mutually influence each other~\cite{xi_material_2018}. 
Here, we show that this coupling gives rise to a deformability-dependent organization of contact forces: soft cells at intermediate self-propulsion exhibit a bimodal force distribution and a characteristic network structure in which weak tip–tip interactions are enclosed by strong side–side interactions. 
Importantly, these features require cell deformability rather than elongated cell shape per se. 
Together, our results reveal a distinct mode of force organization that depends on cell deformability and active motility.

Our results indicate that local curvature at sharp cell tips contributes to setting the strength of weak tip–tip interactions and thereby influences the weak-force peak in the bimodal force distribution.
In living cells, cell-contour curvature is physically constrained by the mechanics of the plasma membrane and its underlying cortex and is further regulated by curvature-sensing and curvature-generating mechanisms~\cite{peter_bar_2004, carman_bar_2018}.
These mechanisms may limit how sharply cell contours can bend, consistent with the curvature constraint examined in our model.
Our results therefore suggest that biological limits on local curvature may tune the strength of weak cell–cell interactions and thereby influence force distributions within multicellular assemblies.

The packing geometry of highly deformed cells provides a possible explanation for why weak tip–tip contacts tend to cluster rather than being dispersed throughout the system. 
The spatial clustering of multiple sharp tips may provide an energetically favorable way to accommodate these poorly space-filling regions, while allowing the remaining cell surfaces to form extended side–side contacts.
Such a packing mechanism could account for the spontaneous organization of weak tip–tip interactions within loops of strong side–side interactions, which manifests as a distinct off-diagonal cluster in the persistence diagram. 
To our knowledge, such an off-diagonal cluster has not been reported in persistence diagrams of force networks in granular materials~\cite{kramar_persistence_2013} and was absent in our simulations of non-deformable particles.

The accumulation of sharp cell tips near the center of the force-loop motif identified by topological data analysis is reminiscent of multicellular rosettes observed in living tissues~\cite{blankenship_multicellular_2006, sugimura_vertex_2024}.
Multicellular rosettes are transient multicellular structures that play important roles in morphogenesis. 
In epithelial tissues, they arise as intermediates during neighbor exchange (T1 transition)~\cite{blankenship_multicellular_2006}. 
In mesenchymal cells undergoing mesenchymal–epithelial transition, multicellular rosettes facilitate intercalation into the overlying epithelium~\cite{gredler_multicellular_2023}.
Previous theoretical studies using the Cell Vertex Model have primarily explained rosette formation in epithelial tissues by introducing positive feedback between junction contractility and shrinkage~\cite{perez-verdugo_tension_2023, sknepnek_generating_2023, lange_vertex_2025}, motivated by sequential formation of supracellular actomyosin cables and their contractility~\cite{blankenship_multicellular_2006}.
Other studies have stabilized higher-order vertices by introducing phenomenological rules that delay the resolution of four or higher-order vertex~\cite{das_controlled_2021, erdemci-tandogan_effect_2021}.
In contrast, our simulations showed that rosette-like structures emerge autonomously through mechanical interactions among contacting cells, without explicitly implementing such feedback on junction mechanics or delays in junction/vertex remodeling.
Moreover, rather than forming higher-order vertices in which multiple cell vertices converge, our simulations produce clusters of sharp cell tips.
Because this mechanism relies on cell deformability and motility rather than junction contractility, it may be particularly relevant to tissues in which mature cell–cell junctions are less pronounced and cells are highly motile, such as mesenchymal tissues.
Together, these findings suggest an alternative physical mechanism for the emergence of rosette-like multicellular structures, distinct from the higher-order vertices generated in polygonal cell models.

Although our analysis reveals a deformability-and motility-dependent force network, its functional significance in multicellular systems remains unclear.
One possible function is to maintain structural integrity in compressed tissues.
Cancer cells are softer than their normal counterparts~\cite{alibert_are_2017, xu_cell_2012} and are often subjected to compression due to aberrant proliferation~\cite{linke_compressive_2024}.
A previous study suggested that, despite the absence of a percolated backbone of stiff cells and the presence of a large fraction of unjammed, motile cells, the coexistence of stiff and soft cells helps maintain tissue structure in breast tumors~\cite{fuhs_rigid_2022}.
If the strong-force network identified here contributes to load bearing, our findings suggest a possible mechanism by which assemblies of soft cells alone could resist mechanical loads and thereby maintain tissue integrity under compression, including in tumors.

Our findings demonstrate that cell deformation is not merely a passive response to applied forces, but can itself shape how forces are organized across multicellular system.
Topological data analysis enables us to uncover such force-network structures that are not apparent from spatially-averaged force distributions alone.
Future studies combining network-level topological analysis with experiments will help reveal how self-organized mechanical structures emerge in active cell assemblies.

\clearpage
\subsection*{Author contributions}
Conceptualization: K.Y., K.S.
Methodology: K.Y.
Software: K.Y.
Formal analysis: K.Y.
Investigation: K.Y.
Writing – original draft: K.Y., K.S, N.S.
Writing – review \& editing: K.Y., K.S., N.S., S.I.
Supervision: K.S., N.S., S.I.
Funding acquisition: K.S., N.S., S.I.

\subsection*{Data availability}
\sloppy
The data are available from the corresponding authors upon reasonable request. 
The code used for the numerical simulations can be downloaded from \url{https://github.com/Sugimuralab/FCCforce}.

\fussy
\subsection*{Acknowledgments}
The authors thank Arthur Vidal for discussions during this project.
This study was financially supported by JSPS KAKENHI Grants 25H01364 to K.S. and N.S., 25K07242 to N.S., and 24H01931 and 26H00951 to S.I.
This study was also supported by the JSPS Core-to-Core Program “Advanced core-to-core network for the physics of self-organizing active matter (JPJSCCA20230002).

\subsection*{Declaration of Interests}
The authors declare no competing interests.

\clearpage

\bibliographystyle{unsrt}

\bibliography{main.bib}

\end{document}